\RequirePackage[2018-12-01]{latexrelease}
\documentclass{gGAF2e}

\usepackage{amsmath}
\usepackage{aas_macros}

\usepackage{nohyperref}
\usepackage{graphicx}	
\usepackage[dvipsnames]{xcolor}
\usepackage{newtxtext,newtxmath}
\usepackage[T1]{fontenc}
\usepackage{adjustbox}
\usepackage{upgreek}
\usepackage{ulem,cancel}
\usepackage{nicefrac, xfrac}

\usepackage{academicons}
\usepackage{ae,aecompl}
\usepackage{bm}
\usepackage[utf8]{inputenc}
\usepackage{amsfonts}
\usepackage{amssymb}

\definecolor{midblue}{rgb}{0.0,0.4,0.7}
\definecolor{mypurple}{rgb}{0.7,0.3,0.8}

\definecolor{PineGreen}{HTML}{008B72}
\definecolor{Berry}{HTML}{FF2052}

\newcommand{\dd}{\mathrm{d}}        
\newcommand\sound{_\text{s}} 
\newcommand\deriv[2]{\frac{\partial#1}{\partial#2}}
\newcommand\mean[1]{\langle #1\rangle}
\newcommand\meanh[1]{\langle #1\rangle_{xy}}
\renewcommand\vec[1]{\bm{#1}}
\newcommand\kin{_\text{k}} 
\newcommand\magn{_\text{m}} 

\newcommand{\betacr}{\beta_\text{cr}}
\newcommand{\betagas}{\beta_\text{m}}
\newcommand{\ecr}{\epsilon_\text{cr}}
\newcommand{\ecri}{\epsilon_{\rm{cr0}}}

\newcommand\bh{\tilde{b}}
\newcommand\uh{\tilde{u}}

\DeclareMathAlphabet{\mathsc}{OT1}{cmr}{m}{sc}
\def\testbx{bx}%
\DeclareRobustCommand{\ion}[2]{%
\relax\ifmmode
\ifx\testbx\f@series
{\mathbf{#1\,\mathsc{#2}}}\else
{\mathrm{#1\,\mathsc{#2}}}\fi
\else\textup{#1\,{\mdseries\textsc{#2}}}%
\fi}

\newcommand{\cm}{\,{\rm cm}}    
\newcommand{\km}{\,{\rm km}}    
\newcommand{\p}{\,{\rm pc}}     
\newcommand{\kpc}{\,{\rm kpc}}  
\newcommand{\g}{\,{\rm g}}      
\newcommand{\s}{\,{\rm s}}      
\newcommand{\Myr}{\,{\rm Myr}} 
\newcommand{\Gyr}{\,{\rm Gyr}}  
\newcommand{\kms}{\km\s^{-1}}    
\newcommand{\mkG}{\,\upmu{\rm G}} 
\newcommand{\erg}{\,{\rm erg}}  

\begin{document}

\jvol{00} \jnum{00} \jyear{2012} 

\markboth{Qazi et. al.}{{Magnetic buoyancy instability in galaxies}}


\title{{\textit{Mathematical Aspects of Geophysical and Astrophysical Fluid Dynamics}}: Magnetic buoyancy instability in galaxies}

\author{Yasin Qazi$^{\dag\ast}$\thanks{$^\ast$Corresponding author. Email: Y.Qazi@newcastle.ac.uk},
Anvar Shukurov${^\dag}$,
Devika Tharakkal${^\ddag}$,
Frederick A.~Gent$^{\S\P\dag}$\\\vspace{6pt}
${^\dag}$School of Mathematics, Statistics and Physics, Newcastle University, Newcastle upon Tyne, NE1 7RU, UK\\
${^\ddag}$Department of Physics, University of Helsinki, PO Box 64, FI-00014, Helsinki, Finland\\
${^\S}$Nordita, KTH Royal Institute of Technology and Stockholm University, Hannes Alfv\'ens v\"ag 12, Stockholm, SE-106, Sweden\\
${^\P}$HPCLab, Department of Computer Science, Aalto University, PO Box 15400, FI-00076, Espoo, Finland\\
}
\maketitle

\begin{abstract}
We study the nonlinear evolution of the magnetic buoyancy instability in
rotating and non-rotating gas layers (with emphasis on the parameter range
typical of spiral galaxies) using numerical solutions of non-ideal, isothermal
MHD equations.  The unstable magnetic field is either imposed through the
boundary conditions or generated by an imposed $\alpha$-effect. In the case of
an imposed field, we solve for the deviations from the background state which
satisfy periodic boundary conditions. We also include cosmic rays as a weightless
fluid which exerts a dynamically significant pressure and somewhat amplifies
magnetic buoyancy. This version of the instability is known as the Parker
instability. Without rotation, systems with an imposed magnetic field evolve to a
state with a very weak magnetic field, very different from the marginally
stable eigenfunction, where the gas layer eventually becomes very thin as it is
supported by the thermal and turbulent pressures alone. However, this does not
happen when the magnetic field is maintained by the $\alpha$-effect. Rotation
fundamentally changes the development of the instability. A rotating system
develops nonlinear oscillations, and the magnetic field direction changes even
with an imposed magnetic field. We demonstrate that this is caused by the
secondary $\alpha$-effect at large altitudes as the gas flow produced by the
instability becomes helical. The secondary $\alpha$-effect has an anomalous sign
with the $\alpha$-coefficient being negative in the northern hemisphere, whereas
the Coriolis force produces a positive $\alpha$. The mean-field dynamo action
outside the original gas layer can also lead to a switch in the magnetic field
parity from quadrupolar (typical of the mean-field dynamo action in a thin
layer) to dipolar. Altogether, the magnetic buoyancy instability and the
mean-field dynamo action become separated as distinct physical effects in a
nonlinear rotating system. We show that none of the assumptions used in
analytic studies of the Parker instability is corroborated by numerical
results.


\end{abstract}

\section{Introduction}

The magnetic buoyancy (or magnetic Rayleigh--Taylor) instability (MBI)
\citep{N61}, modified and enhanced by cosmic rays, is known as the Parker
instability \citep{Parker1958,P1966,P79}.  The instability is expected to
develop in any magnetised, strongly stratified astrophysical object at a
relatively short time scale \citep[section 2.8.2 of][and references
therein]{SS21}.  A horizontal magnetic field in a gas layer confined by gravity
can become unstable to undular modes \citep{HC87}, which grow exponentially on
a time scale comparable to the sound or Alfv\'en crossing time over the gas
density scale height. For the warm interstellar gas in the Solar neighbourhood,
where the observed scale height is approximately $0.5\kpc$ and both the sound
and Alfv\'en speeds are about $10\kms$, this time scale is about $5\times10^7$
years. This is much shorter than the lifetime of a galaxy. Various mechanisms
have been discussed which could suppress the instability of galactic discs,
including the effects of cosmic ray diffusion \citep{KP83,K87,HZ18} and
rotation \citep{ZK75,FT94,FT95}. However, the MBI is unlikely to be entirely
suppressed in spiral galaxies. Therefore, it is essential to examine its
nonlinear states to understand why the gas distributions observed in spiral
galaxies are not disrupted by this instability.

Aside from its effect on the vertical distributions of the interstellar gas,
magnetic field and cosmic rays, the MBI plays a significant role in the
evolution of galaxies. It contributes to driving galactic outflows (winds and
fountains) contributing to the star formation feedback in an evolving galaxy
\citep{NO17}.

The MBI is active in the Sun \citep[][and references therein]{H85,P14} and
stars.  Because of its importance in galaxies, the instability is extensively
studied in the presence of cosmic rays (the Parker instability), relativistic
particles which produce significant pressure but do not add noticeably to the
weight of the gas. The linear stage of the Parker instability has been
thoroughly studied and the dispersion relation has been obtained for a wide
range of physical models and parameter regimes \citep[e.g.][and references
therein]{GS1993,FT94,FT95,KHR1997,SS21}. However, the nonlinear state of the
MBI is much less understood, in particular because it can only be studied
numerically \citep{KRJH2001}.

Rotation is known to reduce the growth rate of the weak perturbations, but it
does not suppress the instability completely \citep{ZK75,FT94,FT95,KoHaOt2003}.
However, rotation introduces a fundamentally new feature to the system: gas
flows associated with the instability become helical and can drive a mean-field
dynamo that generates a large-scale magnetic field
\citep{P1992,1998Han&Les,MSS1999,The00a,The00b,HKO-ML04,B2005,JoLe08}. A
striking feature is the possibility of quasi-periodic magnetic field reversals
in a rotating system \citep{JoLe08,LHB2008,Machida2013} which we discuss below
and attribute to the dynamo action driven by the MBI. These nonlinear effects
rely on rotation and are especially surprising considering that neither the MBI
nor the mean-field dynamo in a thin layer is oscillatory on its own.

We review the studies of the nonlinear states of the MBI, with and without
cosmic rays, and the magnetic fields either imposed, i.e.\ maintained by an
external source \citep{DT2022a,DT2022b}, or supported by the mean-field dynamo
within the system \citep{QSTGB23,QSTGB2025}.  The text is structured as
follows: the general model setup is outlined in section~\ref{sec:model_setup},
we then discuss the nonlinear states unaffected by rotation in
section~\ref{sec:no_rot}, the effects of rotation are the subject of
section~\ref{sec:rot}, and we provide a summary in
section~\ref{sec:summary}.

\section{Simulations of the nonlinear magnetic buoyancy instability}\label{sec:model_setup}

We solve numerically the non-ideal, compressible, isothermal MHD equations for
the gas density $\rho$, its velocity $\vec{u}$, total pressure $P$ (which
includes the thermal, magnetic, and cosmic-ray contributions), magnetic field
$\vec{B} = \nabla \times \vec{A}$, its vector potential $\vec{A}$ (with the
advective gauge $\phi = \eta\nabla \cdot \vec{A}$), and where appropriate the
energy density and flux of cosmic rays, $\epsilon_\text{cr}$ and $\vec{F}$. We
use the sixth-order in space and third-order in time finite-difference
\textsc{Pencil Code} \citep{brandenburg2002,Pencil-JOSS}.

Most simulations of the MBI instability do not include the dynamo action as the
source of the unstable magnetic field. Therefore, an equilibrium state
introduced as an initial condition is rapidly destroyed by the instability in
such simulations. This has prevented \citet{RSSBF2016} from analysing strongly
nonlinear states of the system. Alternatively, boundary conditions for the
magnetic field can be used to impose a background state maintained throughout
the simulation. However, this would constrain unphysically the evolution of the
system as the fixed boundary conditions would require that the deviations from
the background state vanish at the boundaries.

Therefore, we present here two different approaches to investigating the
nonlinear state of the instability. The first is to derive and solve (fully
nonlinear) equations for the deviations from the background state. In fact,
this is the standard approach to explore the linear MBI (or any other)
instability analytically, but we use it to capture numerically a fully
nonlinear evolution of the perturbations when their magnitude is no longer
small. The boundary conditions for the deviations are not restrictive (we use
periodic boundary conditions in the horizontal planes), so that the
perturbations can evolve freely. In the nonlinear state of the instability, the
magnitude of the deviations from the background state is comparable to that of
the background state, altering it fundamentally; so it is important to make the
model is fully flexible to allow for the possibility of such a strong
modification.

Alternatively, the unstable magnetic field can be maintained by a physical
mechanism is incorporated into the equations solved. The background, unstable
magnetic field in the second group of models discussed here is generated in the
induction equation by an imposed $\alpha$-effect
\begin{equation} \label{eq:alpha}
    \alpha(z)=\alpha_0
    \begin{cases}
    \displaystyle
    \sin \left(\pi z/h_\alpha\right)\,, &|z| \leq h_\alpha/2\,,\\
    \displaystyle
    (z/|z|) \exp \left[-\left(2z/h_\alpha-z/|z|\right)^2\right]\,, &|z|>h_\alpha/2\,,
    \end{cases}
\end{equation}
with a parameter $\alpha_0$ used to control the intensity of the dynamo action. The vertical extent of the dynamo-active layer is $h_\alpha$ on each side of the midplane; the smaller $h_\alpha$, the stronger is the vertical gradient of the magnetic field and the more it is buoyant. In the models discussed, $h_\alpha$ is much smaller than the vertical extent of the computational region. This form ensures that $|\alpha(z)|$ varies smoothly with $z$ near $|z|=h_\alpha$ since discontinuities in $\dd \alpha/\dd z$ strongly affect the dynamo action. The dynamo intensity (in particular, the growth rate of the large-scale magnetic field) depends on the dimensionless number $R_\alpha=\alpha_0 h_\alpha/\eta$. In the models which do not explicitly include rotation (section~\ref{sec:no_rot}), this leads to the $\alpha^2$-dynamo. When differential rotation is included in section~\ref{sec:rot}, the unstable magnetic field is produced by the $\alpha^2\omega$-dynamo. In order to clarify the role of the Coriolis force in the MBI, we neglect it in section~\ref{sec:no_rot} and include rotational effects in full in section~\ref{sec:rot}.
{In a real astrophysical system, the $\alpha$-effect can only emerge because of rotation. Therefore, including it in models which lack overall rotation is not self-consistent but we use the imposed $\alpha$-effect only as a technical tool to generate an unstable magnetic field and our focus is on the consequences of its instability. Moreover, in our study of the interaction between magnetic buoyancy and the imposed dynamo action we feel free to adopt some extreme values of $R_\alpha\ (\gtrsim1)$ which are not likely to be encountered in applications.}

The MHD equations have the standard form; with the notation used here, they are given by \citet{DT2022b} or \citet{QSTGB2025}. In models with an imposed $\alpha$-effect, the term $\alpha\vec{B}$ is included in the induction equation. The initial state is a plane-parallel magneto-hydrostatic equilibrium in the galactic gravitational field, i.e.\ a stratified layer of thermal gas, horizontal magnetic field and,  in some models, cosmic rays.
The governing equations are solved in a rectangular region with the Cartesian coordinates
$(x,y,z)$ corresponding to the cylindrical $(r,\phi,z)$, of the corresponding dimensions $(L_x,\ L_y,\ 2L_z)=(4\text{--}6,\ 4\text{--}12,\ 3\text{--}3.5)\kpc$ designed to accommodate the most rapidly growing MBI mode. The $z$-axis is antiparallel to the gravitational acceleration $\vec{g}(z)$. We consider the forms of $\vec{g}(z)$ typical of galactic discs, varying linearly with $|z|$ near $z=0$ and tending to a constant at $|z|\to\infty$ \citep[e.g.\ equations 17 and 18 of][]{DT2022a}. The boundary conditions are periodic in $x$, sliding periodic in $y$ when differential rotation is included (periodic otherwise). At $z=\pm L_z$, we have $b_x=b_y=0$, $b_z\neq0$ for the magnetic field perturbations in models with an imposed magnetic field and similarly for the total magnetic field in the other models. The gradient of the density (perturbation when the background magnetic field is imposed, and the total density otherwise) is fixed at $z=\pm L_z$ to $\partial\ln\rho/\partial z=\mp h_0^{-1}$ with a scale height $h_0$ equal to that corresponds to the evolving vertical thermal pressure gradient at $|z| = L_z$ or fixed at $h_0=1.5\kpc$.
The boundary conditions for the gas velocity at $|z|=L_z$ are $u_x=u_y=0$ and allow for free gas outflow and restricted inflow \citep[see][for details]{DT2022a}. We allow free escape of cosmic rays at $|z|=L_z$.
The initial conditions represent magneto-hydrostatic equilibrium in an imposed magnetic field of $\vec{B}_0(0)=(3\text{--}7)\mkG\,\widehat{\vec{y}}$ at $z=0$ (its variation with $z$ is obtained from solving the magneto-hydrostatic equilibrium equations)  or a weak Gaussian random field of $10^{-3}\mkG$ in strength at $z=0$ and varying in proportion to $\rho^{1/2}$ in models with imposed $\alpha$-effect (see below). The numerical resolution is $(\Delta x,\ \Delta y,\ \Delta z)=(15,\ 7,\ 13)\p$ or better.

Cosmic rays are described in the
fluid approximation
\citep[e.g,][]{Parker1969,SL1985} where the cosmic ray energy density $\epsilon_\text{cr}$ is governed by
\begin{equation}
    \deriv{\epsilon_\text{cr}}{t} = Q(z)
-\nabla\cdot(\epsilon_\text{cr}\vec{u}) - p_\text{cr}\nabla\cdot\vec{u}
-\nabla \cdot\vec{F}\,,
\end{equation}
in which $\vec{F}$ is the cosmic ray flux, $p_\text{cr} =
\epsilon_\text{cr}(\gamma_\text{cr} - 1)$ is the cosmic ray pressure and $Q(z)$ is the cosmic ray source. The adiabatic index of the ultrarelativistic gas is $\gamma_{\text{cr}}=4/3$ \citep{SL1985}. The cosmic ray flux $\vec{F}$ is introduced in a non-Fickian form, justified and discussed by
\citet{SBMS2005},
\begin{equation}
    \tau_\text{cr} \deriv{F_i}{t} = \kappa_{ij}\nabla_j\epsilon_\text{cr}-F_i\,,
\end{equation}
where $\tau_\text{cr}=10 \Myr$ can be identified with the decorrelation time of the cosmic ray pitch angles. The cosmic ray diffusion tensor $\kappa$ is
\begin{equation}
    \kappa_{ij} = \kappa_{\perp}\updelta_{ij}+(\kappa_{\parallel} -\kappa_{\perp})\hat{B}_i\hat{B}_j,
\end{equation}
where a circumflex denotes a unit vector, $\kappa_\perp= 3.16 \times 10^{25} \, \rm cm^2s^{-1}$
and $\kappa_\parallel= 1.58 \times 10^{28} \, \rm cm^2s^{-1}$ \citep[][and references
therein]{LSFB2015a,Ryu2003}.

In models with an imposed magnetic field, the initial cosmic ray energy density and the initial magnetic field strength and distribution are introduced by specifying the ratios of the cosmic ray and magnetic pressures to the thermal pressure, $\beta_\text{cr}$ and $\beta_\text{m}$. In the background state, these ratios are adopted as constants, but they vary in space and time as the instability develops. Cosmic rays and magnetic fields are introduced in this manner in many analytical studies of the Parker instability. In models with the imposed $\alpha$-effect, the source of cosmic rays is introduced more realistically in
the form
\begin{equation}
    \mathcal{Q}(z) = Q_{0}\exp(-\lvert z\rvert^2/h_\text{cr}^2)\,.
\end{equation}
Supernova explosions are the main sources of cosmic rays in galaxies. A typical supernova (SN) injects about $10^{51}
\erg$ of energy, of which only a few percent are converted into  cosmic rays
\citep[e.g.][]{KOG72,Sc02}.
The scale height of the energy injection is $h_\text{cr} = 100\p$
\citep{vdBT91} and $Q_{0} = 9.4 \times 10^{49} \erg \kpc^{-3}
\Myr^{-1}$ \citep{vdBergh90,vdBT91}.

Details of the simulations discussed can be found in
\citet{DT2022a,DT2022b} for the imposed background magnetic field and \cite{QSTGB23,QSTGB2025} for the unstable magnetic fields produced by the dynamo. Weakly nonlinear states of the MBI in a similar model with the background magnetic field specified as an initial condition are discussed by \citet{RSSBF2016}.

\section{Non-rotating systems} \label{sec:no_rot}

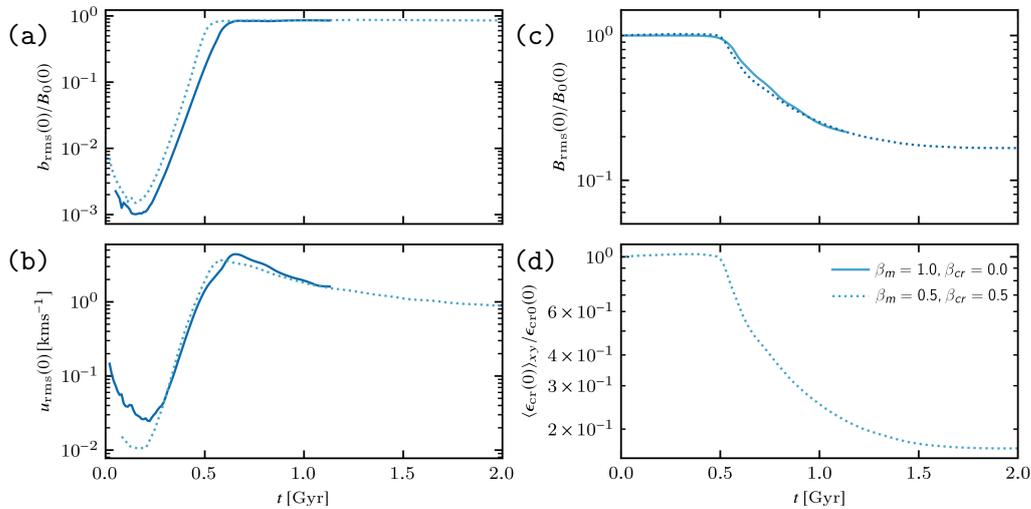
\begin{figure}
\centering
\includegraphics[trim=0cm 0.0cm 0cm 0.0cm, clip=True, width=0.85\columnwidth]{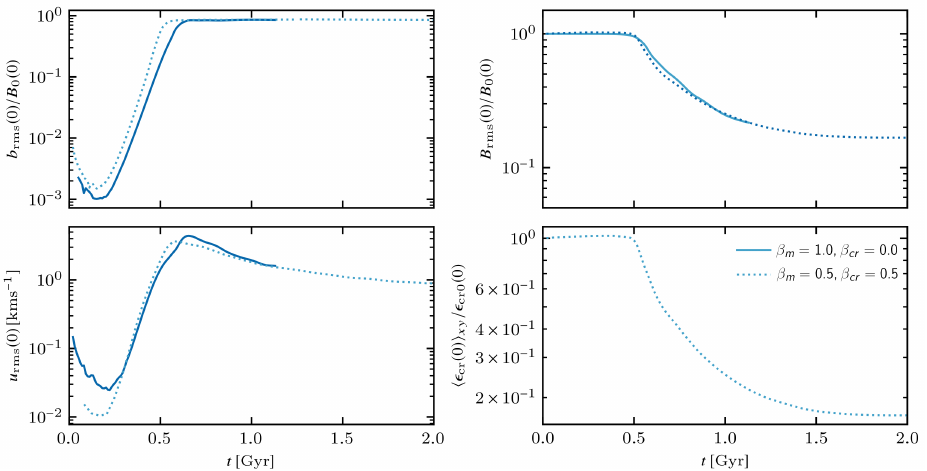}
 \begin{picture}(1,1)
 \put(-390,175){\texttt{(a)}}
 \put(-200,175){\texttt{(c)}}
 \put(-390,90){\texttt{(b)}}
 \put(-200,90){\texttt{(d)}}
 \end{picture}
\caption{The MBI of an imposed magnetic field in a non-rotating system \citep{DT2022a}: the evolution of the root-mean-square magnitudes at the midplane $z=0$ of \textbf{(a)}~the magnetic field perturbation $|\vec{b}|$ normalised to the background strength at $z=0$ and \textbf{(b)}~gas speed in the models with $(\beta_m,\beta_{cr}) = (1,0)$ (solid, no cosmic rays) and $(\beta_\text{m},\beta_\text{cr}) = (0.5,0.5)$ (dotted, equal initial contribution of magnetic field and cosmic rays to the total pressure). The linear stage of the instability ends at about $t\simeq0.4\Gyr$. Panels~\textbf{(c)} and \textbf{(d)} show the similarly normalised total magnetic and cosmic ray energy densities at $z=0$,
$\langle B(x,y,0,t)/B_0(0)\rangle_{xy}$ and $\langle\ecr(x,y,0,t)/\ecri(0) \rangle_{xy}$ respectively, where $\langle\cdots\rangle_{xy}$ denotes the horizontal averaging.
} \label{fig:kg89rmGR}
\end{figure}

The evolution of the root-mean-square (rms) velocity and magnetic fields, shown
in figure~\ref{fig:kg89rmGR} in the case of an imposed magnetic field, reveals
three distinct stages in the development of the instability. In the linear
phase, marked by the exponential growth of magnetic and velocity perturbations
for $t\lesssim 0.4\Gyr$, initial perturbations become dominated by the leading
eigenmode as shown in figures~\ref{fig:kg89rmGR}(a,b). During this phase, the
total magnetic and cosmic ray energy densities remain nearly constant, as shown
in figure~\ref{fig:kg89rmGR}(c), because the perturbations are still weak.
Consistent with earlier analytical and numerical studies \citep[e.g.][]{GS1993,
Ryu2003, LSFB2015a}, cosmic rays enhance the instability. The growth rate
$\Gamma$ of the rms velocity and magnetic fields is lower for models without
cosmic rays  (see {solid} lines in figure~\ref{fig:kg89rmGR}), increasing by
approximately $20\%$ from $19\Gyr^{-1}$ to $25\Gyr^{-1}$ when cosmic rays
contribute pressure equal to that of the magnetic field.

Following the linear and a short transitional (weakly nonlinear) stages at
$0.4<t<0.5\Gyr$, the growth slows down and the system moves toward a
statistical steady state by about $1.6\Gyr$.  During this period, the total
magnetic and cosmic ray energy densities decay, ultimately retaining only a few
percent of their initial values. In contrast, the thermal pressure remains
nearly constant throughout the simulation, decreasing by only about $4\%$ in
the nonlinear regime.

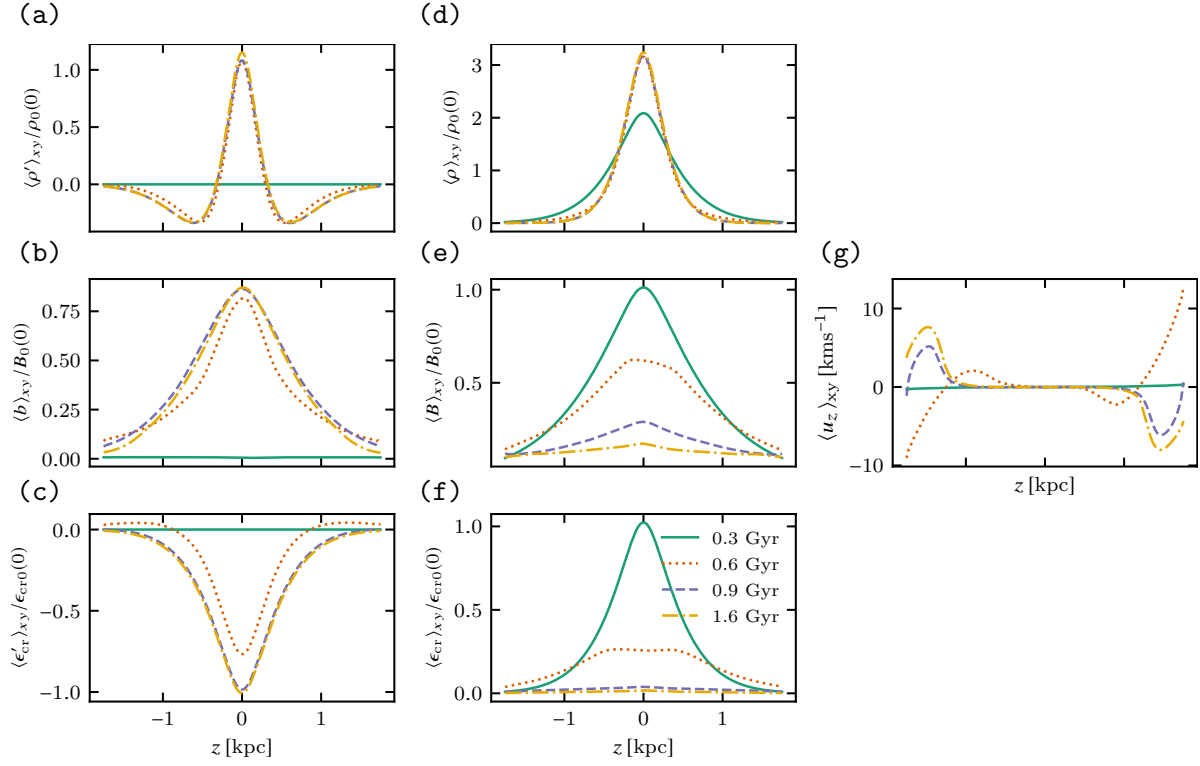
\begin{figure*}
\vspace{3mm}
\centering
\includegraphics[width=1.0\columnwidth]{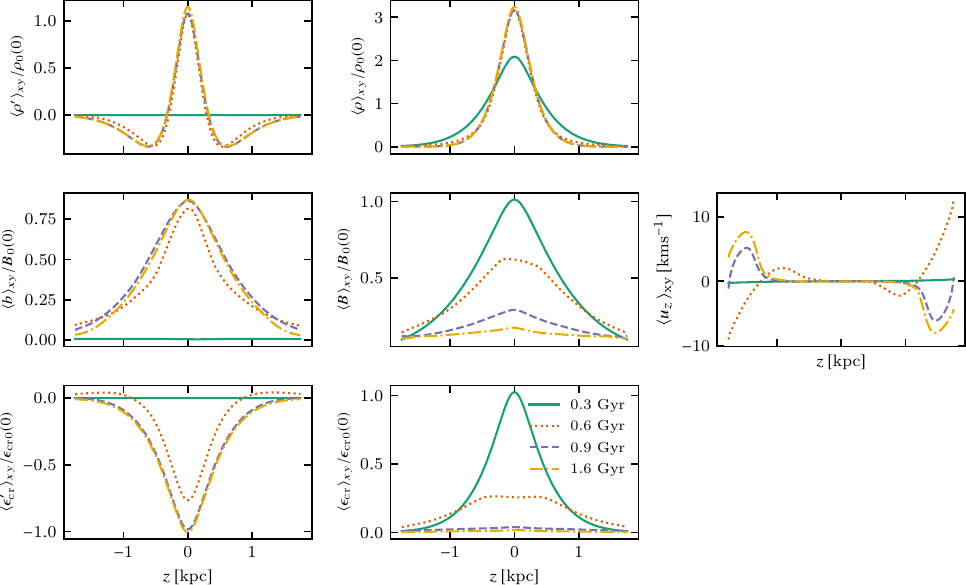}
  \begin{picture}(0,0.)(0,0)
    \put(-220,290){{\texttt{(a)}}}
    \put(-70,290){{\texttt{(d)}}}
    \put(-220,200){{\texttt{(b)}}}
    \put(-70,200){{\texttt{(e)}}}
    \put(80,200){{\texttt{(g)}}}
    \put(-220,110){{\texttt{(c)}}}
    \put(-70,110){{\texttt{(f)}}}
  \end{picture}
	\caption{
	The MBI of an imposed magnetic field in a non-rotating system with $\betacr=0.5$ and $\betagas=0.5$ \citep{DT2022a}: the horizontally averaged vertical profiles of perturbations (left
	column, {(a)}--{(c)}) and total profiles
	(middle  column, {(d)}--{(f)}).
	The solid, dotted, dashed and dash-dotted lines show consecutive evolutionary stages (at times shown in the legend of panel~{(f)}). Panel \textbf{(g)} shows the horizontally averaged vertical velocity normalised to the sound speed $\meanh{u_z}/c\sound$.
}
\label{fig:kg89_profiles}
\end{figure*}

The spatial redistribution of the gas, magnetic field and cosmic rays is
illustrated in figure~\ref{fig:kg89_profiles}, which presents the horizontally
averaged deviations from the background distributions  in panels (a)--(c) and
(g) as well as the distributions in panels (d)--(f) of the total gas density,
magnetic field and cosmic rays, respectively, in the linear, transitional, and
nonlinear phases of the instability. In the linear stage, the perturbations,
periodic in horizontal planes, have vanishing horizontal averages.
Non-vanishing horizontal averages arise only due to nonlinear effects. The
perturbation in the gas density in the transitional and nonlinear stages is
positive near the midplane and negative away from the disc: the nonlinear
instability leads to a reduction in the density scale height and the gas disc
becomes thinner.  As a result, the mean gas density at the midplane increases
from $7\times10^{-25}\g\cm^{-3}$ to $1.08\times10^{-24}\g\cm^{-3}$ as the
instability develops. The average energy densities of the total magnetic field
and cosmic rays at the midplane are reduced by more than $75\%$, expanding
their vertical profiles. Similar behaviour occurs in the simulations of
\citet[][e.g.\  their figure~10]{Heintz2019} which capture the weakly
nonlinear, transitional stage of the instability.

Figure~\ref{fig:kg89_profiles}(g) shows the planar averages of the vertical velocity $\langle u_z\rangle_{xy}$ at
different times. The nonlinear effects drive a systematic inflow at
$|z|\lesssim 1\kpc$ and an outflow at $|z|\gtrsim 1\kpc$ in the transitional stage
which is transformed into a weaker inflow at $t=0.9\Gyr$ while the
system still adjusts towards the steady state.
At $t=1.6\Gyr$, the system reaches a statistical steady state with a residual inflow with $|\meanh{u_z}|/c_{\text{s}}\simeq0.5$ at $|z|\gtrsim1\kpc$ (which, however, carries little mass).
The magnetic field and cosmic ray energy density continue to decrease, saturating at the midplane values  $ \meanh{B(0)}/B_0(0)\approx0.16$ and  $ \meanh{\ecr(0)}/\ecri(0)\approx0.02$, while the gas density increases to $\meanh{\rho(0)}/\rho_0(0)\approx1.6$.

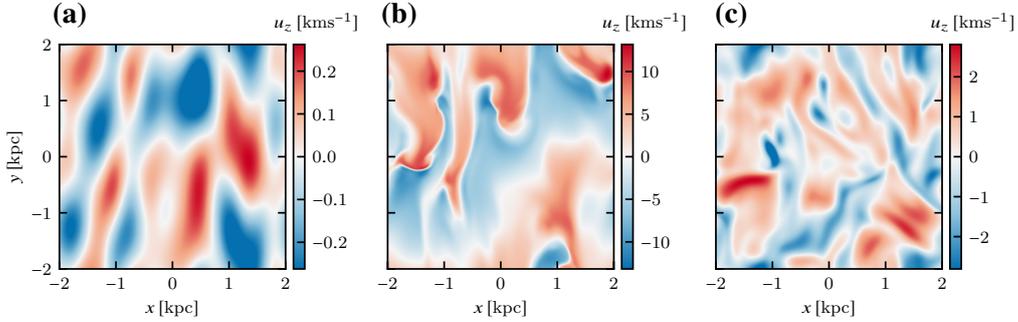
\begin{figure*}
\centering
\includegraphics[trim=0cm 0.0cm 0cm 0cm, clip=True, width=0.85\columnwidth]{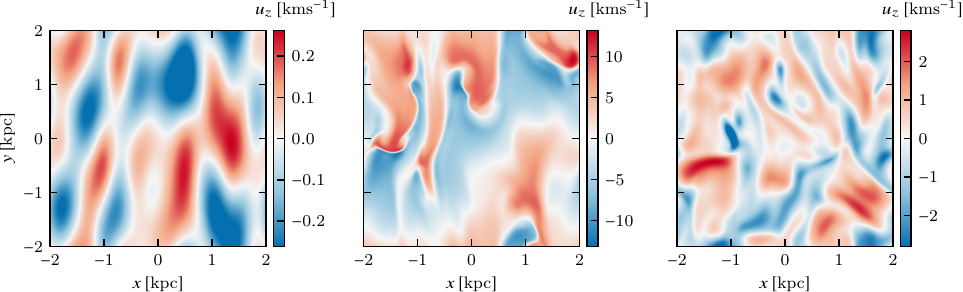}
  \begin{picture}(2,0.)(0,0)
    \put(-363,110.52){{\sf\bf{(a)}}}
    \put(-240,110.52){{\sf\bf{(b)}}}
    \put(-115,110.52){{\sf\bf{(c)}}}
  \end{picture}
	\caption{The vertical velocity perturbation $u_z$ from a model with an imposed magnetic field \citep{DT2022a} with $\betacr=0.5$ and $\betagas=0.5$ at $z=0.4\kpc$ during \textbf{(a)}~the linear stage of the instability ($t=0.3\Gyr$), \textbf{(b)}~the transitional stage ($t=0.6\Gyr$) and \textbf{(c)}~the nonlinear state ($t=0.9\Gyr$).
}
\label{fig:mbi_xy_pert}
\end{figure*}

The magnetic buoyancy instability is driven by the vertical gradient of the magnetic field strength, and it might be expected that it would saturate via reducing the gradient to a marginal value. Instead, the system follows a much more dramatic path, removing the magnetic field altogether. As the instability produces strong vertical magnetic perturbations, the cosmic rays are channelled out from the disc and diffuse along the magnetic field at a high rate. The instability results in a wide spread of both magnetic field and cosmic rays enveloping a relatively thin thermal gas disc. A similar kind of evolution also occurs in the simulations of \citet{Heintz2019} and \citet{GPPS22}.

Figure~\ref{fig:mbi_xy_pert} shows how the vertical velocity pattern evolves. In the linear stage, panel (a), perturbations are regular and average to zero horizontally. During the transitional phase, panel (b), a transient inflow emerges, while outflows appear at higher altitudes (see figure~\ref{fig:kg89_profiles}(g)). By the nonlinear stage, panel (c), the flow becomes chaotic. Likewise, the regular magnetic loops seen in the linear phase break down into a disordered structure, similar to the early nonlinear evolution observed by \citet{LSFB2015a}.

\begin{figure*}
    \centering
    \includegraphics[width=\textwidth]{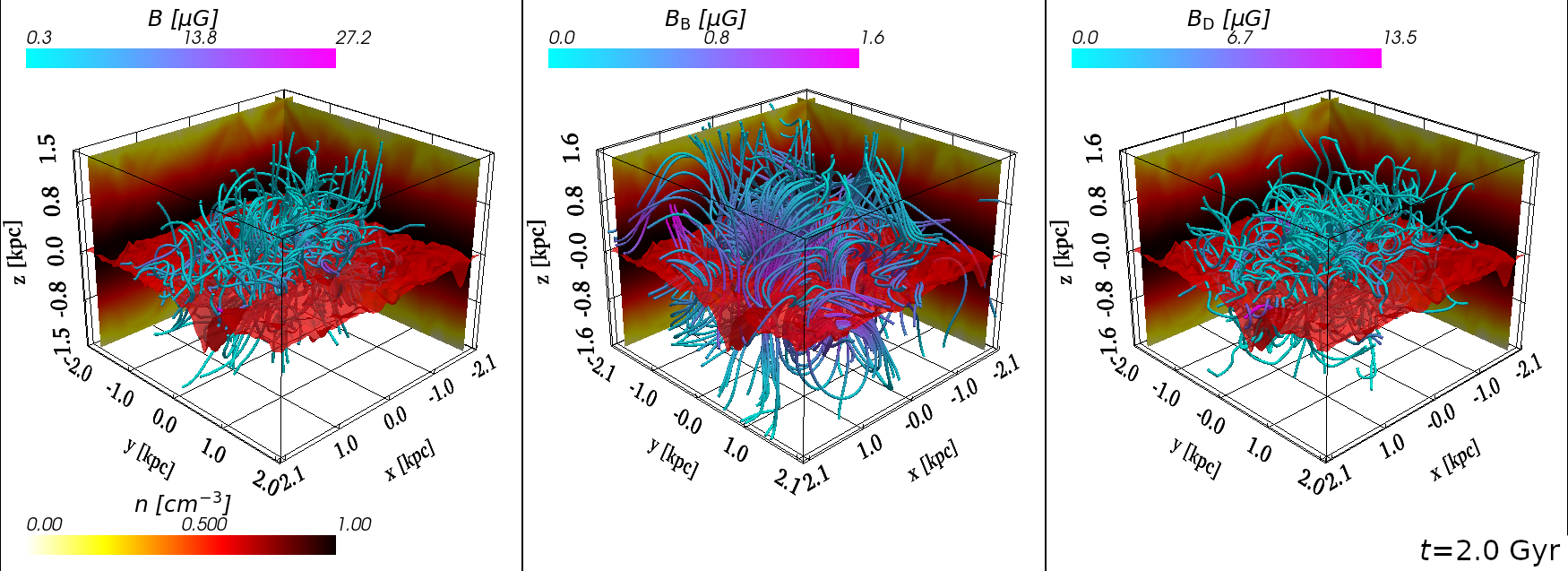}
    \caption{The MBI of a magnetic field generated by a strong imposed
$\alpha$-effect with the overall rotation neglected \citep{QSTGB23}: magnetic
lines  of the total field $\vec{B}$ (left-hand column), separated using
equation~\eqref{GS} with $\ell=200\p$ into contributions at the larger scales
characteristic of the magnetic buoyancy $\vec{B}_\text{B}$ (middle) and the
smaller scales of the imposed dynamo $\vec{B}_\text{D}$ (right-hand column),
with the gas density shown with colour on the vertical planes and as an
isosurface at $z=0$.
    \label{fig:streamlines}
}
\end{figure*}

These results are echoed by similar simulations with an imposed dynamo.
Although the dynamo and magnetic buoyancy instabilities can be distinguished
clearly during their linear stages, at the later stages when the Lorentz force
becomes dynamically significant the effects of the dynamo action and magnetic
buoyancy are strongly intertwined. It is, therefore, harder to identify the
role of each individual instability in the nonlinear behaviour. The symbiosis
of the two instabilities has unexpected results.

To illustrate the three-dimensional magnetic field structure in a
well-developed nonlinear stage, figure~\ref{fig:streamlines} shows the evolving
three-dimensional structure of magnetic lines at various scales in a model with
an imposed $\alpha$-effect. With  $R_\alpha=10$, the scale of the magnetic
field produced by the dynamo is much smaller than the MBI scale, so that the
two instabilities can easily be distinguished in the linear stage.  The total
magnetic field $\vec{B}$ can be separated into the contributions
$\vec{B}_\text{B}$ of the larger scales (characteristic of the MBI) and of the
smaller scales $\vec{B}_\text{D}$ (driven by the imposed dynamo action).
\citet{QSTGB23,QSTGB2025} apply Gaussian smoothing,
\begin{equation}\label{GS}
\vec{B}_\text{B}(\vec{x},t)= \int_V \vec{B}(\vec{x}',t)\, G_\ell(\vec{x}-\vec{x}')\,\dd^3\vec{x}'\,,
\quad
\vec{B}_\text{D}=\vec{B}-\vec{B}_\text{B}\,,
\end{equation}
where $V$ is the whole domain volume, and the smoothing kernel
$G_\ell(\vec{\xi})=(2\pi\ell^2)^{-3/2}\exp[-|\vec{\xi}|^2/(2\ell^2)]$ with
$\ell=200\p$, the scale of the leading dynamo mode at $R_\alpha=10$. We note
that the small-scale part $\vec{B}_\text{D}$ also contains random magnetic
fields produced by nonlinear effects.

\begin{figure}
    \centering
\includegraphics[width=0.7\textwidth]{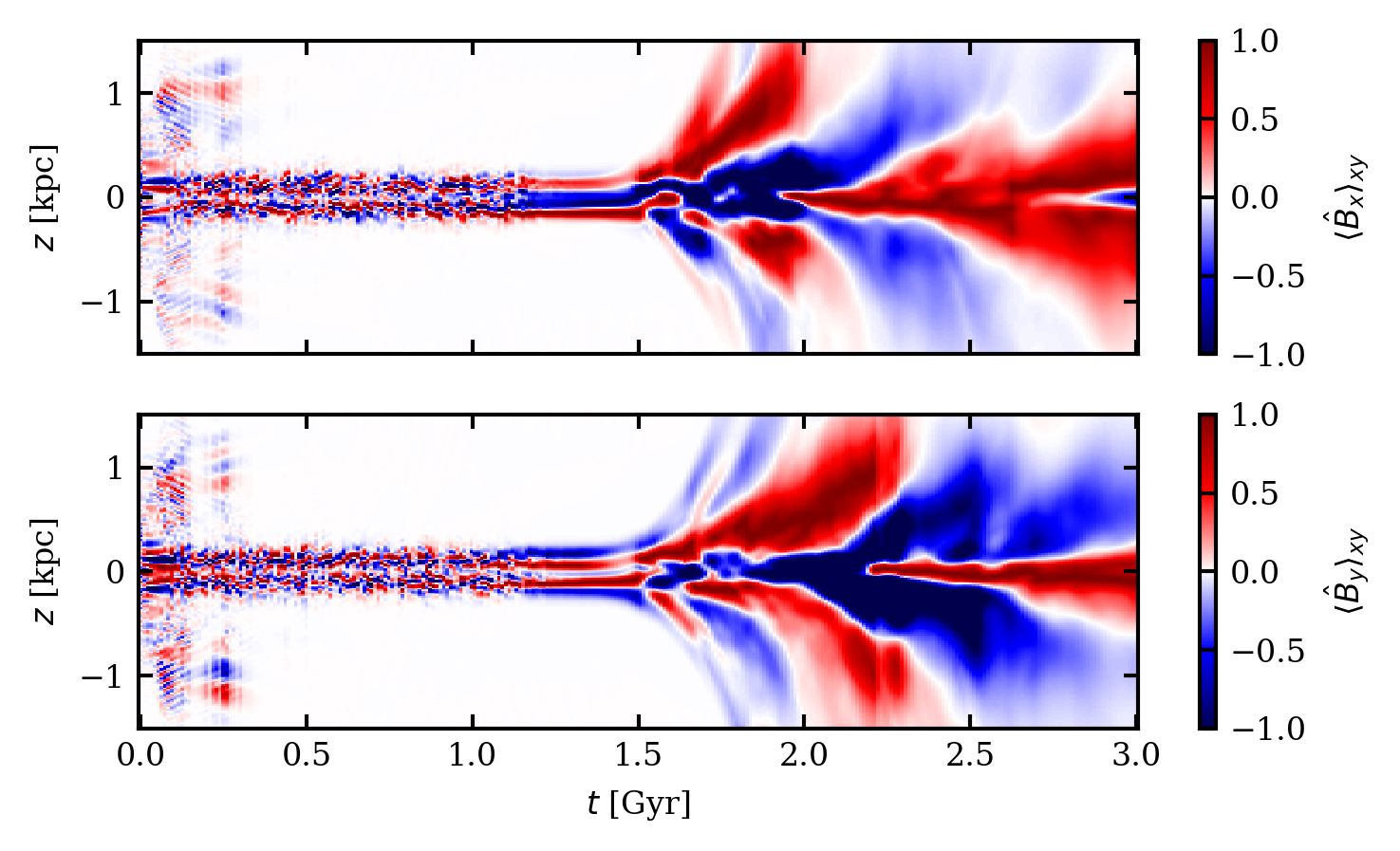}
    \caption{The evolution of the horizontally averaged magnetic field components $\meanh{\widehat{B}_x}$ (upper panel) and $\meanh{\widehat{B}_y}$ (lower panel) for a model with an imposed $\alpha$-effect, normalised to their maximum value at each time \citep{QSTGB23}.
    \label{fig:xy_Bx_By_ha_02_ra_10}
}
\end{figure}

\begin{figure}
    \centering
    \includegraphics[width=0.7\textwidth]{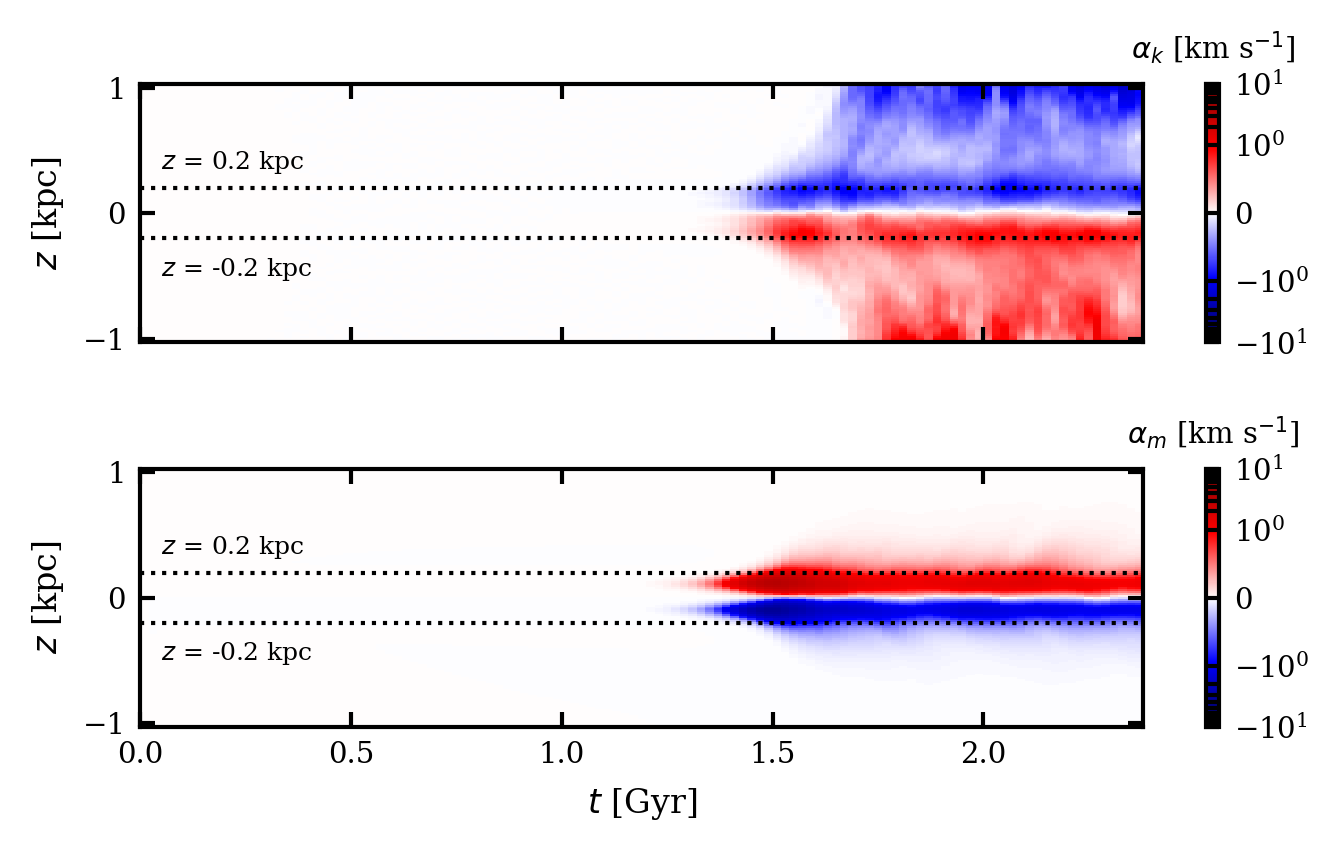}
    \caption{The evolution of the horizontally averaged $\alpha$-coefficients
due to the mean kinetic helicity ($\alpha\kin$, upper panel) and the mean current
helicity ($\alpha\magn$, lower panel), given in equation~\eqref{akm}, in a
model with imposed $\alpha$-effect without rotation \citep{QSTGB23}. The
horizontal dotted lines are shown at $|z|=h_\alpha$.
}
\label{ak_am1}
\end{figure}

Figure~\ref{fig:xy_Bx_By_ha_02_ra_10} illustrates an unexpected feature of the
nonlinear interaction of the MBI and the $\alpha$-effect dynamo: the magnetic
field, which grows monotonically at early times, develops oscillations at
$t\gtrsim1.5\Gyr$ when it becomes strong enough to make the system essentially
nonlinear. This figure shows the evolution of the horizontally averaged magnetic
field components $\meanh{B_x}$ and $\meanh{B_y}$ from a model with a
dynamo-generated magnetic field, normalised to their maximum magnitude at each
time to better expose the field structure at early times when it is still weak.
The magnetic field generated by the kinematic (linear) dynamo is confined to a
relatively thin layer $|z|\lesssim h_\alpha$ and grows monotonically. However,
it spreads to larger altitudes because of the buoyancy (to achieve the scale
height of order $1\kpc$). When fully nonlinear, the magnetic field becomes
oscillatory, reversing its direction at intervals of order $0.5\Gyr$. These
changes in the large-scale magnetic field structure start near the midplane and
are spread by the magnetic buoyancy to larger altitudes.

The oscillations leading to the reversals are explained by the secondary
mean-field dynamo action of the gas flows produced by the MBI. The model of
\citet{QSTGB23}, which includes an imposed $\alpha$-effect at $|z|\lesssim
h_\alpha$, does not include overall rotation and, hence, the Coriolis force.
However, the magnetic field generated by the imposed $\alpha$-effect is helical, and the Lorentz force drives helical motions.
{As a result, the $\alpha$-coefficient of the mean-field dynamo equation is composed of kinetic and magnetic contributions, expressed \citep[e.g.][section~7.11.2 of]{SS21} as:
\begin{equation}\label{alpha}
\alpha = \alpha\kin + \alpha\magn,,
\end{equation}
where, using horizontal averages,
\begin{equation}\label{akm}
\alpha\kin = -\tfrac{1}{3} \tau_0 \meanh{\vec{\uh} \cdot (\nabla \times \vec{\uh})},, \qquad
\alpha\magn = \tfrac{1}{3} \tau_0 \frac{\meanh{\vec{\bh} \cdot (\nabla \times \vec{\bh})}}{4\pi\rho},,
\end{equation}
and $\tau_0$ is the correlation time of the perturbed flow and $\vec{\uh}$ and $\vec{\bh}$ are the deviations from the horizontal averages $\meanh{B}$ and $\meanh{U}$ for the magnetic field and velocity, respectively.
}
Since the helicity of the
magnetic field is opposite to that of the imposed $\alpha$, the resulting
secondary $\alpha$-effect has the sign opposite to the imposed one.
Figure~\ref{ak_am1} shows the contributions to the $\alpha$-coefficient from
the gas flows and the current helicity, derived using equation~\eqref{akm} in
this model. 
This figure
confirms that $\alpha\kin$ is antisymmetric in $z$ and $\alpha\kin<0$ at $z>0$,
the sign opposite to that attributable to the Coriolis force, and the sign of
the imposed $\alpha$.
To confirm this interpretation,
\citet[][section~4.1]{QSTGB23} propose a one-dimensional nonlinear mean-field
dynamo model which includes equations for the horizontal magnetic field
components, the vertical component of the Navier--Stokes equation for the
vertical flow due to magnetic buoyancy and the $\alpha$-coefficient as expressed by equation~\eqref{eq:alpha}.
The model is remarkably successful in reproducing the nonlinear oscillations
observed in the three-dimensional simulations.
{The secondary $\alpha$-effect leads to even more dramatic changes when the overall rotation is included explicitly in Section~\ref{sec:rot}.}

The oscillations in a system driven by the $\alpha$-effect emerge only when the
system becomes nonlinear. In isolation, both the $\alpha$-effect dynamo in a
thin layer and the MBI are non-oscillatory. As we demonstrate in
section~\ref{sec:rot}, such nonlinear oscillations are a generic property of
the MBI in a rotating system, and they are related to the secondary mean-field
dynamo action driven by the MBI.

\section{Magnetic buoyancy in rotating systems}\label{sec:rot}

\begin{figure*}
\centering
\includegraphics[width=0.9\textwidth]{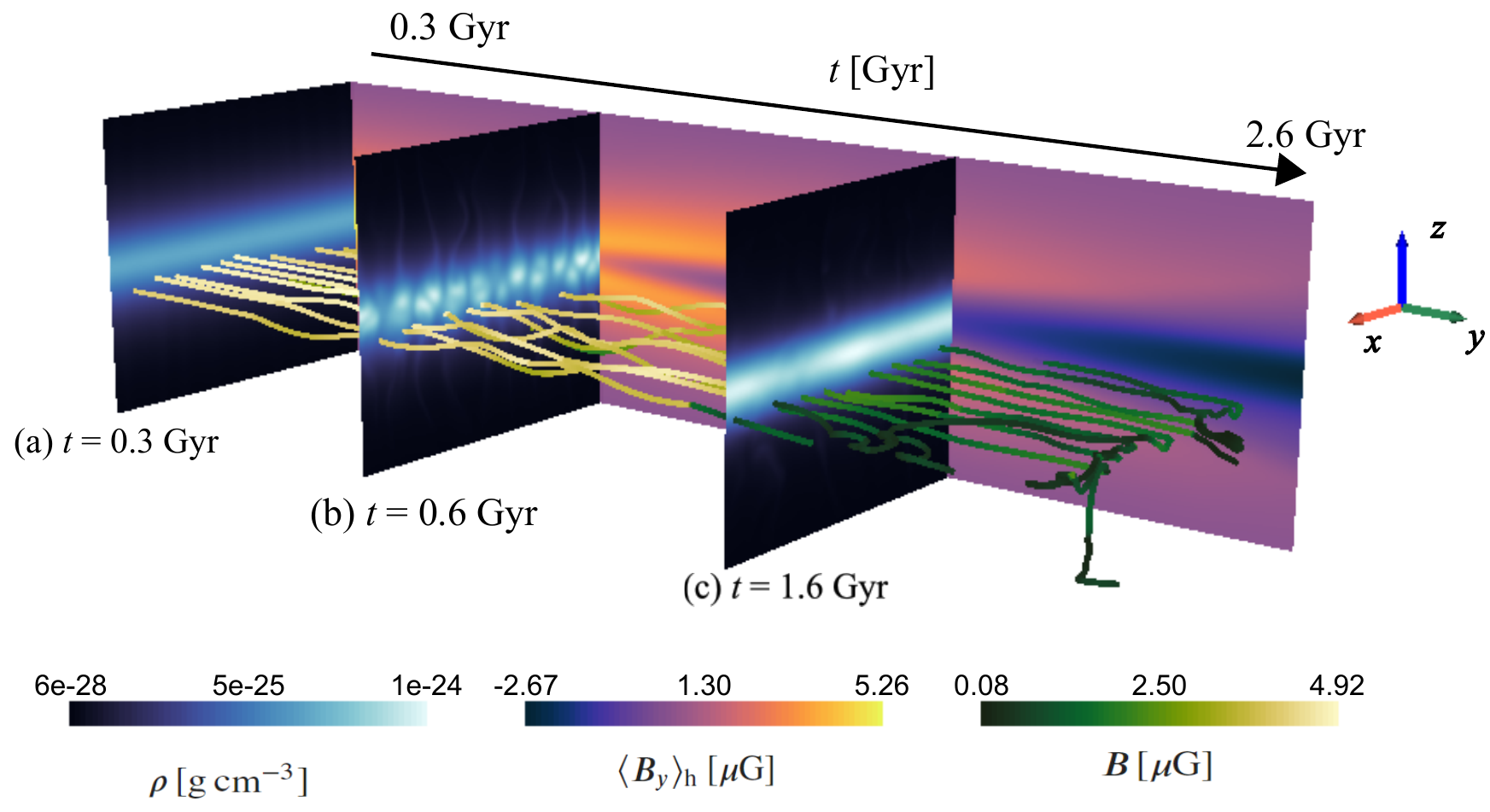}
\caption{The evolution of the gas density and magnetic field in a model of a rotating system with an imposed magnetic field \citep{DT2022b} is illustrated for its three significant epochs: \textbf{(a)}~the linear stage, \textbf{(b)}~beginning of the magnetic field reversal in the early nonlinear stage and  \textbf{(c)}~the advanced nonlinear state (the specific simulation times are indicated for each frame). Selections of magnetic lines are shown with colour representing the local magnetic field strength. The horizontal average of the azimuthal magnetic field $\langle B_y\rangle_{xy}$ (denoted $\langle B_y \rangle_\text{h}$ near the colour bar) is shown with colour on the vertical $(z,t)$-plane as it evolves continuously (rather than at discrete times used for the magnetic lines). The gas density distribution is shown with colour on the vertical $(x,z)$-planes.}
\label{fig:3d_vi_simb}
\end{figure*}

Rotation changes the system fundamentally. In this section we include
differential rotation with the angular velocity $\Omega(r)$ and shear $S=r\,
\dd \Omega/\dd r<0$ (we recall that $x$ is the analogue of the cylindrical
radius $r$). A rotating system with an imposed magnetic field, in contrast
to the non-rotating system, retains a strong magnetic field near the midplane as the MBI develops and, unexpectedly, the direction of the mean magnetic
field can be reversed as the perturbations become stronger than the imposed
field and have, on average, the opposite direction. As discussed below, this is
a manifestation of nonlinear oscillations similar to those discussed above.
Moreover, for a sufficiently strong rotation, the parity of the mean magnetic
field can change from quadrupolar at early stages of the evolution to dipolar
in the developed nonlinear stage.

\subsection{Magnetic field reversals and nonlinear oscillations}\label{NO}

Figure~\ref{fig:3d_vi_simb} presents a pictorial summary of the changes in the magnetic field and gas density as the instability develops through its linear stage and then saturates in a model with an imposed magnetic field and rotation \citep{DT2022b}. During the linear phase, at $t=0.3\Gyr$, the total magnetic field and gas density retain the structure of the imposed fields with weak perturbations in the gas density. By the weakly nonlinear stage at $t=0.6\Gyr$, both the gas density and magnetic field are strongly perturbed to the extent that the mean azimuthal magnetic field $\meanh{B_y}$ starts reversing (as can be seen in the distribution of  $\langle B_y\rangle_{xy}$ shown in the vertical plane of the figure). The reversal is complete in the late nonlinear stage at $t=1.6\Gyr$.

The reversal of the magnetic field in the models with imposed field is especially unexpected because the background magnetic field $B_y>0$ has been maintained throughout the simulations. These simulations were not extended further to verify if further reversals will follow but
{it appears plausible} that the system with the reversed magnetic field can become unstable and the reversal cycle is repeated, so that \citet{DT2022b} observe an early stage of a nonlinear quasi-periodic oscillation similar to that in the system with imposed $\alpha$-effect discussed above.

\begin{figure*}
\centering
\includegraphics[width=\textwidth]{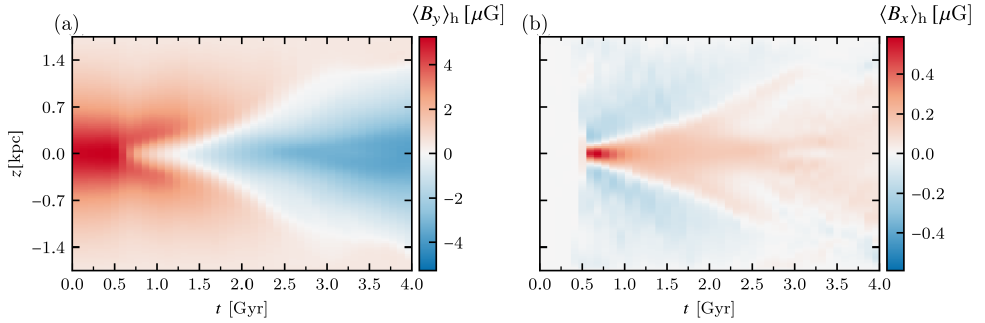}
\caption{The evolution of the horizontally averaged magnetic field components, $\meanh{B_y}$ (left-hand column) and $\meanh{B_x}$ (right-hand column) in a model with
rotation and imposed magnetic field $B_y$ \citep{DT2022b}. The horizontally averaged azimuthal field $\langle {B_y}\rangle_{xy}$ decreases after $t=0.6\Gyr$, and undergoes a reversal in sign at $t\approx 1.6\Gyr$,  with the reversal then spreading to higher altitudes.     Meanwhile, the mean radial field $\langle {B_x}\rangle_{xy}$ becomes positive and relatively strong near $z=0$ rather abruptly at $t\approx0.5\Gyr$ and then also spreads away from the midplane.}
\label{fig:bxbyxy}
\end{figure*}

The reversal of the magnetic field in this model is clearly visible in figure~\ref{fig:bxbyxy} which presents the horizontal averages of the magnetic field components. The reversal starts in the weakly nonlinear phase at $t=0.5\Gyr$ with a rather abrupt emergence of a relatively strong positive radial magnetic field near the midplane, $\langle {B_x}\rangle_{xy}>0$. The velocity shear with $S<0$ stretches the positive radial field into a negative azimuthal magnetic field, so that $\langle {B_y}\rangle_{xy}$ starts decreasing and reverses at $t=1.6\Gyr$ (figure~\ref{fig:bxbyxy}(a)). The total horizontal magnetic field strength $(\langle {B_x}\rangle_{xy}^2+\langle {B_y}\rangle_{xy}^2)^{1/2}$ decreases to a minimum before increasing again, as $\meanh{B_y}$ decreases to zero and then re-emerges with the opposite direction.
\citet{DT2022b} show that the reversal of the magnetic field is explained by the mean-filed dynamo action by the gas flows produced by the MBI and identify the correlator $\langle b_z\,\partial u_x/\partial z\rangle$ as the driver of the reversal.

\begin{figure}
\centering
\includegraphics[width=\columnwidth]{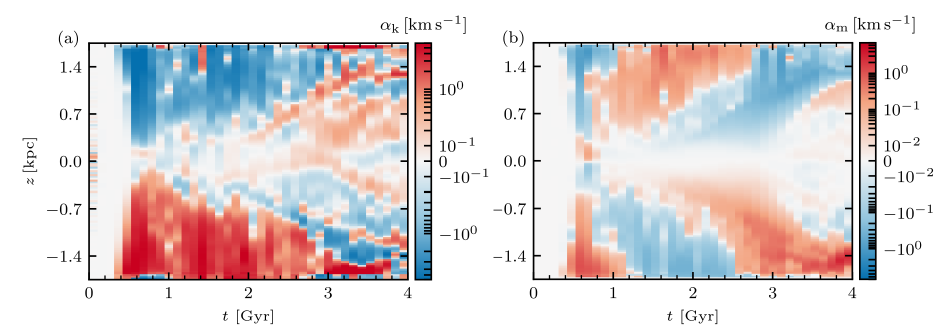}
\caption{The evolution of \textbf{(a)}~the kinetic helicity $\alpha\kin$ and \textbf{(b)}~magnetic contribution to the helicity $\alpha\magn$, given in equation~\eqref{akm}, in the model of figure~\ref{fig:bxbyxy}
}
\label{fig:alpha}
\end{figure}

Gas motions driven by the MBI in a rotating system become helical under the action of the Coriolis force. This results in the $\alpha$-effect which produces a large-scale radial magnetic field. As we argue below, the dynamo action causes the magnetic reversal.
The kinetic and magnetic contributions to the $\alpha$-effect, {equation~\eqref{akm},} in the model of figure~\ref{fig:bxbyxy} are shown in figure~\ref{fig:alpha}.
The spatial structure of $\alpha\kin$ is relatively simple during the early nonlinear phase but becomes increasingly complex as the system evolves. In the later stages, particularly near the midplane, $\alpha\kin$ typically becomes positive above the midplane ($z> 0$) and negative below ($z< 0$), as expected from the action of the Coriolis force \citep[e.g.\ section~7.1 of][]{SS21}. Moreover, the region where $\alpha\kin$ remains predominantly positive (though small in magnitude) expands with time to larger altitudes $\lvert z \rvert$.

As expected, the sign of the current helicity, which produces $\alpha\magn$, is generally opposite to that of $\alpha\kin$ across most values of $z$ and $t$. This leads to a magnetic back-reaction that counteracts the kinetic contribution, ultimately saturating the dynamo and driving the system into a statistical steady state by $t\gtrsim 3\Gyr$.

A negative $\alpha\kin$ at $z>0$ (corresponding to a positive kinetic helicity)
appears to be a characteristic feature of flows driven by magnetic buoyancy or
other magnetically induced instabilities such as the magneto-rotational
instability (MRI). \citet{1998Han&Les}, using a model of reconnecting magnetic
flux ropes, argued that such a sign inversion is plausible in magnetic
buoyancy-driven dynamos. Similarly, \citet{The00a}, in his linear analysis of
the mean electromotive force from the magnetic buoyancy instability in
spherical geometry, found $\alpha<0$ in the unstable region of the northern
hemisphere (analogous to $z>0$ here), although the anomalous sign went largely
unnoticed \citep{The00b}. In contrast, \citet{BrSc98} reported $\alpha\kin> 0$
at $z> 0$ in their analysis of the buoyancy-driven $\alpha$-effect, while
\citet{BrSo02} found $\alpha\kin <0$ in the upper layers of MRI-driven dynamos.
Similar results have emerged from simulations of MRI-driven dynamos by
\citet{DBS24}.

As discussed in section~\ref{sec:no_rot}, simulations with imposed $\alpha$-effect confirm the possibility of a reversal of the magnetic field direction and show that this is a part of nonlinear oscillations. \citet{QSTGB2025} propose a one-dimensional nonlinear mean-field dynamo model which includes vertical flows due to magnetic buoyancy and all three components of the mean magnetic field. The model reproduces the nonlinear oscillations remarkably well both qualitatively and quantitatively.

\begin{figure*}
    \centering
    \includegraphics[width=\textwidth]{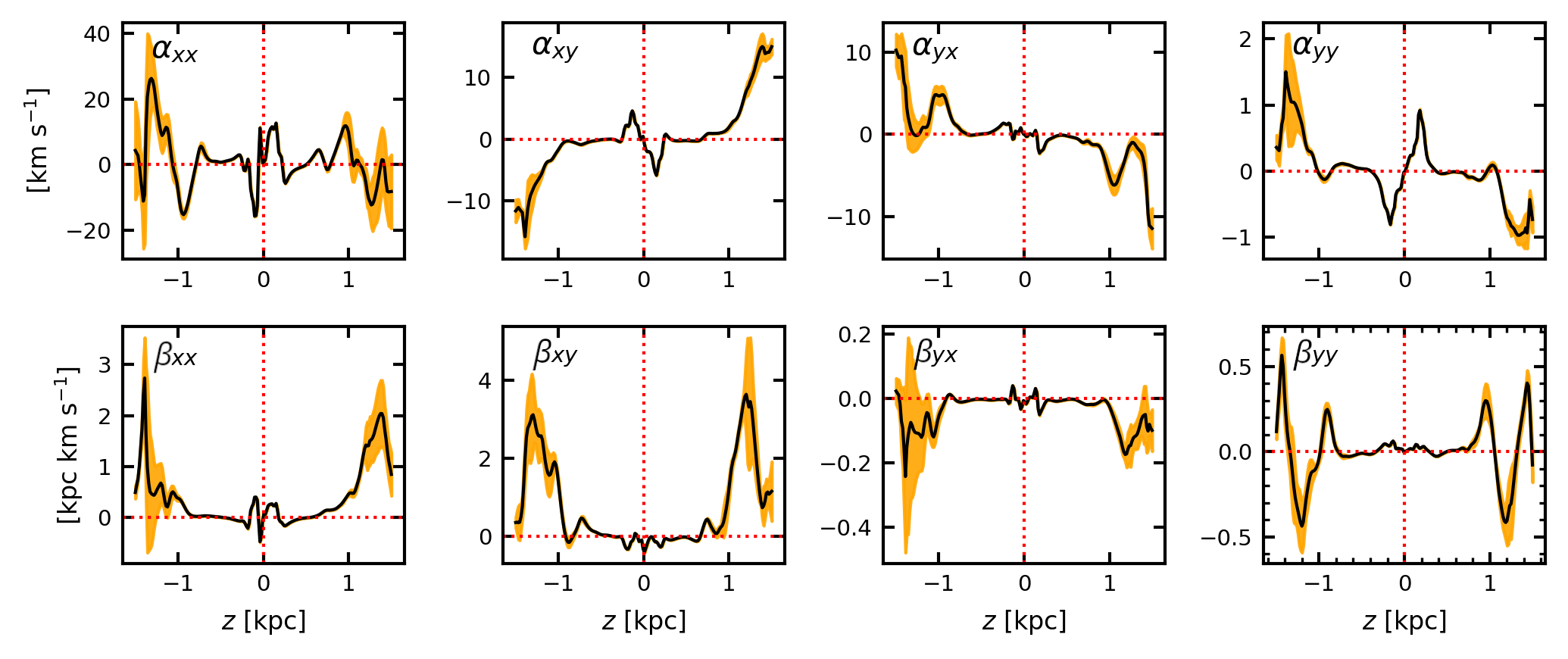}
   \caption{The elements of the turbulent transport tensors introduced in
equation~\eqref{EMF} in the simulations of \citet{QSTGB2025} which include
differential rotation and imposed $\alpha$-effect. The yellow shading
indicates the standard deviations based on bootstrap resampling of the time
series of $\vec{\mathcal{E}}$.}
    \label{fig:IROS}
\end{figure*}

In order to confirm the conclusion that the $\alpha$-effect driven by the MBI has the opposite sign to that produced by the Coriolis force, \citet{QSTGB23,QSTGB2025} derived the elements of the turbulent transport tensors  $\alpha_{ij}$ for the $\alpha$-effect and $\beta_{ij}$ for the  turbulent magnetic diffusivity using the
single value decomposition (SVD) and the iterative removal of sources (IROS) methods introduced by \citet{2015Bendre} and \citet{bendre2023iterative}.
Using horizontal averages of the magnetic field,  the components of the electromotive force $\mathcal{E}_i =
\langle \vec{u}\times\vec{b} \rangle_i$ are approximated by $\mathcal{E}_i = \alpha_{ij}\langle \vec{B_j}\rangle -
\beta_{ij}(\nabla\times \langle \vec{B} \rangle)_j$.
Explicitly,
\begin{align}\label{EMF}
\begin{pmatrix} \mathcal{E}_x\\ \mathcal{E}_y\end{pmatrix}
= \begin{pmatrix} \alpha_{xx} &\alpha_{xy}\\ \alpha_{yx} &\alpha_{yy} \end{pmatrix}
\begin{pmatrix} \mean{B}_x \\ \mean{B}_y \end{pmatrix}
&-\begin{pmatrix} \beta_{xx} &\beta_{xy}\\ \beta_{yx} &\beta_{yy} \end{pmatrix}
\begin{pmatrix} (\nabla\times\mean{\vec{B}})_x \\ (\nabla\times\mean{\vec{B}})_y \end{pmatrix}.
\end{align}
The tensors $\alpha_{ij}$ and $\beta_{ij}$ are assumed to be independent of time, which is valid during the saturated, statistical steady state of the system. Because of the horizontal averaging, the results only depend on the vertical coordinate $z$, and only the horizontal magnetic field components are considered because the horizontal average of the vertical magnetic field vanishes due to periodic boundary conditions. The diagonal elements of $\alpha_{ij}$ represent the scalar $\alpha$-effect, while the off-diagonal components (notably $\alpha_{xy}$) describe vertical transport of mean magnetic fields. The diagonal components of $\beta_{ij}$ represent turbulent magnetic diffusion. When the magnetic diffusivity varies in space, the mean magnetic field is transported against its gradient \citep[turbulent diamagnetism, e.g.\ section~7.9 of][]{SS21}.

Figure~\ref{fig:IROS} shows the elements of the turbulent transport tensors in the non-linear regime of the simulations of \citet{QSTGB2025} which include both differential rotation and the imposed $\alpha$-effect. 
{We note that the imposed $\alpha$ is not captured in these calculations and does not appear in this figure because it is not associated with any velocity and magnetic fields.} Consistently with the results shown in figure~\ref{fig:alpha},
$\alpha_{xx} + \alpha_{yy}$ is large, antisymmetric about the midplane, and predominantly negative for $z>0$. The off-diagonal elements of $\alpha_{ij}$ and $\beta_{xx}$ increasing in magnitude with $|z|$ support the transport of the mean magnetic field towards $z=0$ that counteracts the buoyant escape, aiding the saturation of the MBI.

\subsection{Parity of the magnetic field}\label{PMF}
Models with imposed $\alpha$-effect reveal another striking feature of the nonlinear MBI: a complete change in the magnetic field parity in a deeply nonlinear stage. The mean magnetic field in the models discussed above preserves its quadrupolar parity determined by the symmetry of the imposed magnetic field and/or the $\alpha$-effect dynamos in a thin layer with $\alpha>0$ at $z>0$. In a quadrupolar field, $B_x$ and $B_y$ are symmetric with respect to the midplane at $z=0$, whereas $B_z$ is antisymmetric. A dipolar field has the opposite symmetry, $B_x(-z)=-B_x(z)$, $B_y(-z)=-B_y(z)$ and $B_z(-z)=B_z(z)$.

\begin{figure}
    \centering
    \includegraphics[width=0.7\textwidth]{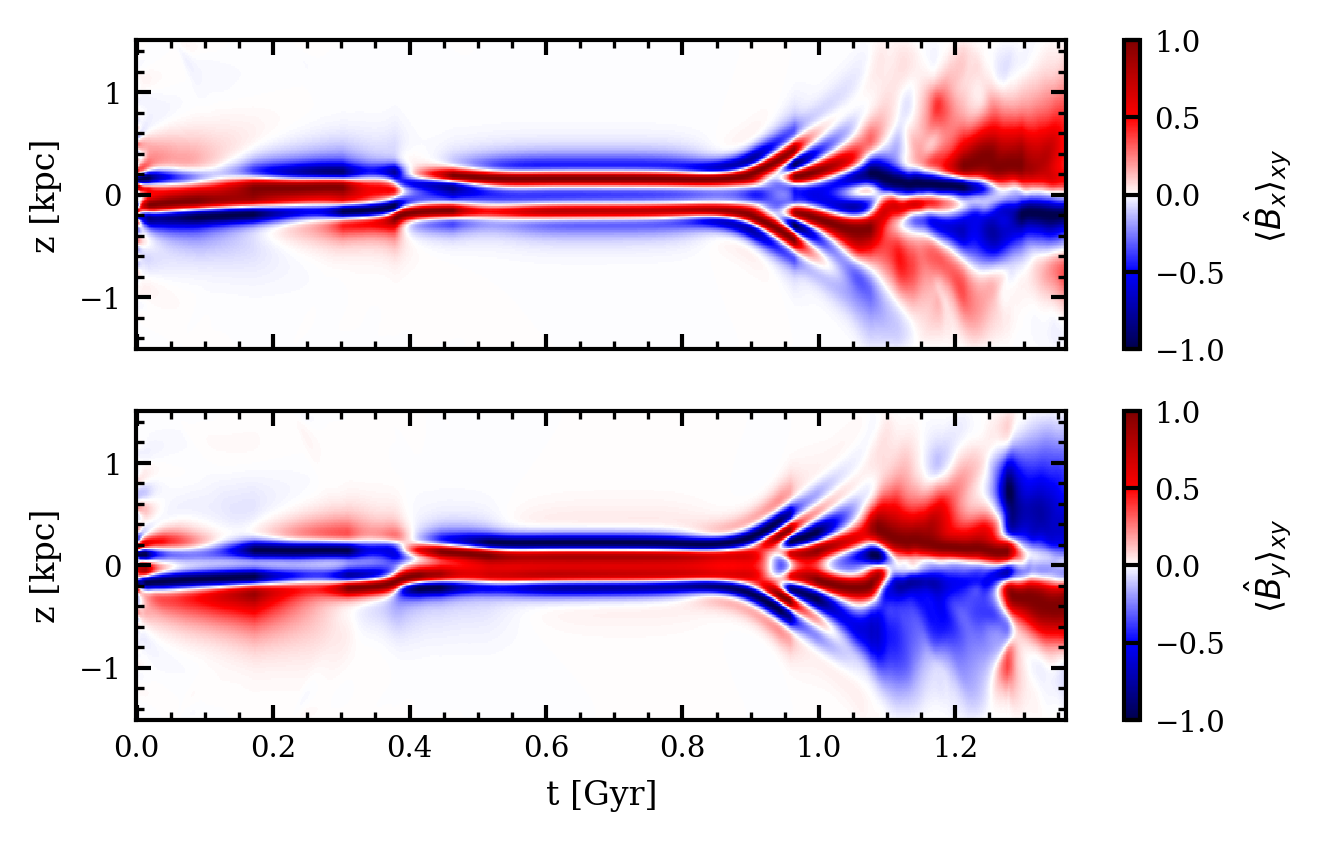}
    \caption{The evolution of the horizontally averaged magnetic field components $\langle\widehat{B}_x\rangle_{xy}$ (upper panel) and $\langle \widehat{B}_y \rangle_{xy}$ (lower panel) in a model with an imposed $\alpha$-effect and rotation  \citep{QSTGB2025}. The hat indicates that each component has been normalised to its maximum magnitude at each time to visualise the magnetic field structure when it is still weak.}
    \label{fig:xy_averages}
\end{figure}

Figure~\ref{fig:xy_averages} illustrates the evolution of the magnetic field parity under the influence of rotation in one of the models explored by \citet{QSTGB2025}. In order to show more clearly the structure of the magnetic field at early times when it is still weak, this figure shows the magnetic field components normalised to their maximum strength at each time, $|\langle\widehat{B}_x\rangle_{xy}|\leq1$ at any $t$ and likewise for $\langle\widehat{B}_y\rangle_{xy}$. During the linear stage, the magnetic field grows monotonically, maintaining its initial quadrupolar symmetry and having reversals at about $t=0.4$ and $0.95\Gyr$ discussed above. However, at $t\gtrsim1.1\Gyr$, the field changes its parity.

As discussed by \citet{QSTGB2025}, the change in parity is due to the secondary dynamo action of the gas flows produced by the MBI which have the anomalous sign of the mean helicity (in comparison with that produced by the Coriolis force), with $\alpha<0$ at $z>0$ (see above).
This interpretation is confirmed using a one-dimensional mean-field dynamo model \citet{QSTGB2025}, which reproduces quantitatively all aspects observed in the simulations discussed here.

In order to change the magnetic parity from that determined by the (imposed) $\alpha$-effect near $|z|=0$, the secondary dynamo action has to be strong enough. Therefore, this regime of the instability is more likely to occur in the central parts of astrophysical discs (of spiral galaxies and in accretion discs) where the velocity shear associated with differential rotation is sufficiently strong to produce vigorous secondary dynamo action.

\section{Summary} \label{sec:summary}

The nonlinear evolution of the MBI is strikingly different from any conceptions based on its linear properties. Rather than simply reducing the magnetic field gradient, the instability in its well-explored form (a unidirectional unstable magnetic field in a non-rotating system) leads to a global weakening of the magnetic field and escape of cosmic rays. In the resulting hydrostatic equilibrium, the gas layer is maintained almost exclusively by the thermal pressure (and turbulent pressure if available), so that it becomes thinner than in the initial state while the remaining weak magnetic field and cosmic rays (if present) have much larger scale heights than the gas.

When the unstable magnetic field is generated by a dynamo, the system develops nonlinear magnetic oscillations, which also occur in rotating systems with an imposed magnetic field. The oscillations do not occur in systems with magnetic buoyancy or dynamo action alone but emerge from their nonlinear coupling. Moreover, the joint action of the dynamo, which produces an unstable horizontal magnetic field, and its buoyancy can lead to a change in the parity of the mean magnetic field from quadrupolar to dipolar if the secondary $\alpha$-effect dynamo driven by the MBI is strong enough. This usually requires a sufficiently strong differential rotation to support the $\alpha\omega$-dynamo.

The diversity and complexity of the nonlinear states of the MBI is due to the fact that gas flows which it produces become helical under the action of the Coriolis and Lorentz forces, and can drive a secondary $\alpha$-effect dynamo. The $\alpha$-effect induced by the Lorentz force has the opposite sign to the conventional $\alpha$-effect driven by rotation (Coriolis force). In a rotating system, the magnetic buoyancy instability and mean-field dynamo action become inseparable, and their synergy leads to nonlinear states that cannot be anticipated on the basis of the linear theories alone.

The efficiency and outcome of the interaction between magnetic buoyancy and dynamo action depend on the gas scale height (the thinner is the disc, the more buoyant is its magnetic field for a fixed field strength) and rotation rate (stronger differential rotation leads to stronger dynamo action), and varies with distance to the centre of an astrophysical disc. Large-scale magnetic fields in galaxies and accretion discs may not simply grow steadily in strength as suggested by the standard mean-field disc dynamo theory. Instead, once the field becomes dynamically significant, it may oscillate in strength and direction. Such complex behaviour is more likely in the central parts of astrophysical discs where both the dynamo action and buoyancy effects are stronger.

\section*{Acknowledgements} \label{sec:ack}
The authors benefited from valuable discussions at the Nordita workshop
‘Towards a Comprehensive Model of the Galactic Magnetic Field’ at Nordita
(Stockholm) in 2023, supported by NordForsk and Royal Astronomical Society. FAG
acknowledges support of the Swedish Research Council (Vetenskapsrådet) grant no.\ 2022–03767.
{We are grateful to anonymous referee for useful comments which have helped to improve the text substantially.}

\bibliographystyle{gGAF}
\bibliography{refs} 
\end{document}